\documentclass[preprint,a4paper]{elsarticle}
\usepackage{mathrsfs}
\usepackage{amsfonts,amsmath,amssymb,amsthm}

\usepackage{latexsym}
\usepackage{tikz}
\usepackage{hyperref}
\usepackage{geometry}

\usepackage{hyperref}
\usepackage{tikz}
\usepackage[all]{xy}
\usepackage[mathscr]{eucal}
\usepackage{soul}
\usepackage{color}

\newtheorem{thm}{Theorem}[section]
\newtheorem{exam}[thm]{Example}
\newtheorem{defn}[thm]{Definition}
\newtheorem{prop}[thm]{Proposition}
\newtheorem{lem}[thm]{Lemma}
\newtheorem{coro}[thm]{Corollary}

\newtheorem{rem}[thm]{Remark}

\newcommand{\DDa}{\mathord{\mbox{\makebox[0pt][l]{\raisebox{-.4ex}{$\downarrow$}}$\downarrow$}}}

\newcommand{\UUa}{\mathord{\mbox{\makebox[0pt][l]{\raisebox{.4ex}{$\uparrow$}}$\uparrow$}}}

\newcommand{\ua}{\mathord{\uparrow}}
\newcommand{\da}{\mathord{\downarrow}}
\newcommand{\rom}[1]{\rm{\uppercase\expandafter{\romannumeral #1}}}

\newcommand{\wt}{\widetilde}

\makeatletter
\def\@author#1{\g@addto@macro\elsauthors{\normalsize%
    \def\baselinestretch{1}%
    \upshape\authorsep#1\unskip\textsuperscript{%
      \ifx\@fnmark\@empty\else\unskip\sep\@fnmark\let\sep=,\fi
      \ifx\@corref\@empty\else\unskip\sep\@corref\let\sep=,\fi
      }%
    \def\authorsep{\unskip,\space}%
    \global\let\@fnmark\@empty
    \global\let\@corref\@empty  
    \global\let\sep\@empty}%
    \@eadauthor={#1}
}
\makeatother

\begin{document}
\begin{frontmatter}
\title{Free dcpo-algebras via directed spaces \tnoteref{t1}}
\tnotetext[t1]{Research supported by NSF of China (Nos. 11871353,12001385).}

\author{Yuxu Chen}
\ead{yuxuchen1210@sina.com}

\author{Hui Kou\corref{cor}}
\ead{kouhui@scu.edu.cn}

\author{Zhenchao Lyu}
\ead{zhenchaolyu@scu.edu.cn}

\cortext[cor]{Corresponding author}
\address{Department of Mathematics, Sichuan University, Chengdu 610064, China}


\begin{abstract}
Directed spaces are natural topological extensions of dcpos in domain theory and form a cartesian closed category. We will show that the D-completion of free algebras over a Scott space $\Sigma L$, on the context of directed spaces, are exactly the free dcpo-algebras over dcpo $L$, which reveals the close connection between directed powerspaces and powerdomains. By this result, we provide a topological representation of upper, lower and convex powerdomains of dcpos uniformly.  

  \vskip 2mm

\end{abstract}

\end{frontmatter}

\section{Introduction}
Powerdomains play an important role in domain theory to provide mathematical models for the denotational semantics of nondeterministic functional programming languages. The upper (Smyth) powerdomains, the lower (Hoare) powerdomains and the convex (Plotkin) powerdomains are the most classical ones and are investigated by many scholars \cite{Hec91,Hec92}. They can be defined through the form of free dcpo-algebras (see \cite{AJ94,CON03}).

The existence of free dcpo-algebras can be proved by the Adjoint Functor Theorem \cite{AJ94}. In \cite{AJ2008}, Jung, Moshier and Vicker gave an algebraic representation of free dcpo-algebras of general dcpos via covering relations. The lower powerdomain of a dcpo $L$ has a simple topological representation $\Gamma(L)$, the lattice of nonempty Scott closed subset of $L$. For continuous dcpos, the upper powerdomain has a simple topological representation $\mathcal{Q}(L)$, the set of nonempty Scott compact saturated subsets of $L$, endowed with the reverse inclusion order. However, there is no topological representation of upper powerdomains and convex powerdomains for general dcpos. This problem was discussed in \cite{Hec91,Hec92} and pointed out after \cite[Theorem 6.2.14]{AJ94} by Abramsky and Jung.

In \cite{Kei12}, the separately continuous algebraic operations on $T_0$ spaces and the laws that they satisfy, both identities and inequalities, were extended to the D-completion. The free algebras were investigated in the context of d-spaces. Directed spaces were introduced in \cite{YK2015}, which are very natural topological extensions of dcpos \cite{CKL2022,XK2022,YK2015}. Directed spaces are equivalent to $T_0$ monotone determined spaces defined by Ern\'e in \cite{ERNE2009}. The category of directed spaces, denoted by $\bf DTop$, is cartesian closed and is a coreflective subcategory of topological spaces.  Analogous to dcpo-algebra, algebras in the category $\bf DTop$, called dtop-algebras, can be defined.  Directed lower, upper, and convex powerspace can be defined in the same way of lower, upper and convex powerdomains as concrete examples of free dtop-algebras. In \cite{XK2020,XK2022}, concrete topological representations of directed lower, upper and convex powerspaces were given. Although the existence of the free dtop-algebras over a directed space was proved by the Adjoint Functor Theorem in \cite{CK2022b}, the proof is not constructive. There is no concrete representation of free dtop-algebras over general directed spaces.

In this paper, we will give the concrete topological representations of free dtop-algebras over any directed space via term algebras, and show that the D-completion of the free dtop-algebra over a Scott space $\Sigma L$ is exactly the free dcpo-algebra over $L$ with respect to the same signature and inequalities, which shows the close connection between directed powerspaces and powerdomains. As applications, we get the concrete topological representation of the convex, lower and upper powerdomains of an arbitrary dcpo $L$ uniformly. 

\section{Preliminaries}

We assume some basic knowledge of domain theory, category theory and topology, as in, e.g., \cite{AJ94,Gou13,CON03}.


Let $P$ be a poset. Given any subset $A$ of $P$, denote by $\ua A$ (resp., $\da A$) the upper (resp., lower) subset of $P$ genrated by $A$. For an element $a$, we write $\ua a$ for $\ua \{a\}$ briefly. Denote by $\sigma(P)$ the Scott topology on $P$ and define $\Sigma P = (P, \sigma(P))$. A topological space is called a Scott space if it is equal to $\Sigma L$ for some dcpo $L$. 

Topological spaces will always be supposed to be $T_0$. For a topological space $X$, its topology is denoted by $\mathcal{O}(X)$ or $\tau$. Let $C(X)$ denote the lattice of closed subsets of $X$, and let $\Gamma(X) = C(X) \backslash \{\emptyset\}$. The partial order $\sqsubseteq$ defined on $X$ by $x\sqsubseteq y \Leftrightarrow  x\in \overline{\{y\}}$
is called the  specialization order, where $\overline{\{y\}}$ is the closure of $\{y\}$. All order-theoretical statements about $T_0$ spaces, such as upper sets, lower sets, directed sets, and so on, always refer to  the specialization order $\sqsubseteq $.\ For any net $\xi = (x_i)_{i \in I}$,\ where $I$ is a directed set,\ we write it $(x_i)$ for short.\ Given any $x\in X$, $(x_i)$ is called converging to $x$,\ denoted by $(x_i)  \rightarrow x $,\ if $(x_i)$ is eventually in every open neighborhood of $x$.\  

\vskip 3mm

A topological space is called a monotone convergence space if it is a dcpo under its specialization order and its topology is coarser than the Scott topology. Monotone convergence spaces are also called d-spaces. Denote by $\bf Top_d$ the category of all d-spaces with continuous maps as morphisms. Given any $T_0$ space $X$, the D-completion of $X$ is defined as follows.

\begin{defn}\rm \cite{Kei09}
For a topological space $X$, the D-completion $\xi_X: X \to X^d$ is defined (up to categorical equivalence) by the universal property that given a continuous $f: X \to Y$ into a monotone convergence space $Y$, there exists a unique continuous function $f^d: X^d \to Y$ such that $f^d \circ \xi_X = f$, i.e., such that the following diagram commutes:

$$\xymatrix{
  X \ar[r]^{\xi_X} \ar[dr]_{\forall f} &  X^d \ar[d]^{\exists ! f^d}    \\
 & Y                    }$$
\end{defn}

A subset $A$ of a poset $P$ is called d-closed if for every directed $ D \subseteq A$ with supremum $\bigvee D$ existing, $\bigvee D \in A$. The set of all complements of d-closed sets of $P$ forms a topology on $P$, called the d-topology (see \cite{Kei09}). Let $cl_d$, called d-closure, denote the closure operator relative to the d-topology. 

For a $T_0$ space $X$, the D-completion of $X$ exists and has a standard representation as follows \cite{Kei12}. Let $cl_d (\Psi(X))$ be the d-closure of $\Psi(X)$ in $\Gamma(X)$, where $\Psi(X) = \{\da x: x \in X\}$ and $\Gamma(X)$ is the set of nonempty closed subsets of $X$, endowed with the inclusion order. Let $\wt{X} = cl_d (\Psi(X))$ and $X^d = (\wt{X},\tau)$, where $\tau$ is the topology inherent from the upper topology on $\Gamma(X)$. In fact, $\tau = \{\Diamond U: U \in \mathcal{O}(X)\}$, where $\Diamond U = \{A \in \wt{X}: A \cap U \not = \emptyset\}$. Define $ \xi_X : X \to X^d$ by $\xi_X(x) = \da x$.

For any d-space $Z$ and any continuous map $f:X \to Z$, the corresponding $f^d$ is defined as follows. $f^d= h \circ g$, where $h$ is the continuous map from $ \Psi(Z)$ to $Z$ with $h(\da x) = x$, and $g: X^d \to \Psi(Z)$ is defined to be $g(A) = cl(f(A))$ for any $A \in X^d$.

\vskip 3mm

In the following, we introduce some basic properties of convergence classes and directed spaces.

For any set $X$, we define $\Phi(X)$ to be the class of all nets in $X$. A convergence class  $\mathcal{C}$ in $X$ is a relation between $\Phi X$ and $X$, i.e., $\mathcal{C}$ is a subclass of $\{ (\xi,x): \xi \text{ is a net in }X,\ x \in X\}$. We write $\xi \Rightarrow_\mathcal{C} x$ iff $(\xi,x) \in \mathcal{C}$.
A topological space is called determined by convergence class $\mathcal{C}$,
denoted by $(X,\mathcal{C}(X))$ or $\mathcal{C}X$ for short, if
 $$U \in \mathcal{C}(X)\ \Longleftrightarrow\   \forall  (\xi,x) \in \mathcal{C},\  x \in U \text{ implies that } \xi \text{ is eventually in } U. $$

Restrict the nets to be directed subsets, we will get the notion of a directed space.

Let $(X,\mathcal{O}(X))$ be a $T_0$ space. Every directed subset $D\subseteq X$ can be  regarded as a monotone net $(d)_{d\in D}$. Set
$DS(X)=\{D\subseteq X: D \ {\rm is \ directed}\}$ to be the family of all directed subsets of $X$. For an $x\in X$, we denote  $D\rightarrow x$  to mean  that $x$ is a limit of $D$, i.e., $D$ converges to $x$ with respect to $X$. It is easy to see that for any $(D,x)\in DS(X)\times X$, $D\rightarrow x$ if and only if $D\cap U\not=\emptyset$ for any open neighborhood of $x$. Set $\mathcal{D}_X =\{(D,x)\in DS(X)\times X:  \   D\rightarrow x \}$ to be the set of all pairs of directed subsets and their limits in  $X$. The topology on $X$ determined by $\mathcal{D}_X$ is denoted by $\mathcal{D}(X)$, called the directed-topology on $X$, i.e., a subset $U\subseteq X$ is directed-open iff for all $(D,x)\in DLim(X)$, $x\in U$ implies $D\cap U\not=\emptyset$. Obviously, every open set of $X$ is directed-open, i.e., $\mathcal{O}(X)\subseteq \mathcal{D}(X)$.

\begin{defn}\rm \cite{YK2015} A  topological space $X$ is said to be a directed space if it is $T_0$ and  every directed-open set is open; equivalently, $ \mathcal{D}(X)=\mathcal{O}(X)$.
\end{defn}

In $T_{0}$ topological spaces, the notion of a directed space is equivalent to the monotone determined space defined by Ern\'e \cite{ERNE2009}.\  Given any space $X$, we denote by $\mathcal{D}X$ the topological space $(X,\mathcal{D}(X))$. The following are some basic properties of directed spaces.

\begin{thm}\rm \cite{YK2015} \label{convergence}
 Let $X$ be a $T_0$ topological space.
\begin{enumerate}
\item[(1)] For all $U\in \mathcal{D}(X)$, $U=\ua U$.
\item[(2)] $X$ equipped with $\mathcal{D}(X)$ is a $T_0$ topological space such that  $\sqsubseteq_d =\sqsubseteq$, where $\sqsubseteq_d$ is the specialization order relative to $\mathcal{D}(X)$.
\item[(3)] For a directed subset $D$ of $X$, $D\rightarrow x$ iff $D\rightarrow_d x$ for all $x\in X$, where $D\rightarrow_d x$ means that $D$ converges to $x$ with respect to the topology $\mathcal{D}(X)$.
\item[(4)] $\mathcal{D}X$ is a directed space.
\end{enumerate}
\end{thm}

Every dcpo endowed with the Scott topology is a directed space. Many properties of dcpos can be extended to directed spaces (see \cite{FK2017,L2016,XK2022,YK2015}). Denote the category of directed spaces with continuous maps as morphism by $\bf DTop$. Then, $\bf DTop$ is cartesian closed \cite{YK2015,L2016}. Moreover, it is a coreflective full subcategory of $\bf Top$, the category of $T_0$ topological spaces. Given any continuous map $f:X \to Y$ between two topological spaces. Let $\mathcal{D}f: \mathcal{D}X \to \mathcal{D}Y$ be the same set map of $f$. Then $\mathcal{D}$ is the coreflector from $\bf Top$ to $\bf DTop$.

For any two directed spaces $X,Y$, we denote by $X \otimes Y$ the categorical product of $X,Y$ in $\bf DTop$. Given directed spaces $X_i, 1 \leq i \leq n$, denote by $\bigotimes_{i=1}^n D_i$ the categorical product of all $D_i$ in $\bf DTop$, and denote by $\prod_{i=1}^n D_i$ the topological product of all $D_i$. Then $\bigotimes_{i=1}^n D_i = \mathcal{D}(\prod_{i=1}^n D_i)$.

\begin{lem}\rm \cite{CK2022b,YK2015}  \label{dir sep con}
Let $Y$ be a directed space and $\{X_i\}_{1 \leq k \leq n}$ be a family of finite number of directed spaces and $f$ be a map from $\bigotimes_{1\leq k\leq n} X_k$ to $Y$. Then $f$ is continuous iff it is separately continuous, i.e., continuous at each argument.
\end{lem}

In \cite{ZHA2021}, it was shown that every directed space has a Scott completion as follows.

\begin{defn}\rm \cite{ZHA2021}
A Scott completion $(Y,f)$ of a space $X$ is a Scott space $Y$ together with a
continuous function $f: X \to Y$ such that for any Scott space $Z$ and continuous function
$g : X \to Z$, there exists a unique continuous map $\widetilde{g}$ satisfying $g = \widetilde{g} \circ f$.
\end{defn}

Recall that for a directed space $X$, $\wt{X}$ is the carrier set of $X^d$, $\wt{X}$ is a dcpo and $\xi_X(x) = \da x$.

\begin{thm}\rm \cite{ZHA2021} \label{Scott completion}
Let $X$ be a directed space. Then $(\widetilde{X}, \xi_X)$ is a Scott completion of $X$. For any Scott space $Z$ and continuous map $f: X \to Z$, $f = f^d \circ \xi_X$.
\end{thm}

Endowed with the Scott topology, $\wt{X}$ together with $\xi_X$ is the Scott completion of $X$. From now on, we also view a dcpo $L$ as the Scott space $\Sigma L$ if there is no specific explanation. The following statement shows that for a directed space, its D-completion and Scott completion coincide.

\begin{prop}\rm \label{D-completion Scott}
Let $X$ be a directed space. Then, $X^d$ is the Scott completion of $X$.
\end{prop}
\begin{proof}
We need only to show that the topology of $X^d$ is the Scott topology. Since $X^d$ is a d-space, its topology is coarser than the Scott topology. Given any Scott open subset $\mathcal{V}$ of $\wt{X}$ and $A \in \mathcal{V}$, since $A \in \wt{X}$, $A \in cl_{\sigma}\{\da a: a \in A\}$. It follows that $\{\da a : a \in A\} \cap \mathcal{V} \not = \emptyset$. Then, there exists some $a \in A$ such that $\da a \in \mathcal{V}$. That is, $\eta(a) \in \mathcal{V}$. Then, $a \in \eta^{-1}(\mathcal{V})$. Thus, there exits some open subset $U$ of $X$ such that $a \in U \subseteq \eta^{-1}(\mathcal{V})$. Since $\eta(U) = \Diamond U = \{A \in \Gamma(X): A \cap U \not = \emptyset\}$, $A \in \Diamond U \subseteq \mathcal{V}$. Then, the topology of $X^d$ is equal to the Scott topology.
\end{proof}

Let $P$ be a dcpo and $A$ be a subset of $P$. Then the $d$-closure of $A$ is in fact the least sub-dcpo of $P$ that contains $A$. It can also be gained by transfinite induction. Let
$S^{0} = A$. For any ordinal $\alpha$, define  
$$S^{\alpha + 1} = \{x : x = \bigvee D, D \text{ is a directed subset of } S_{\alpha} \}. $$
For any limit ordinal $\beta$, define 
$S^\beta = \bigcup_{\alpha < \beta} S^\alpha$. Let 
$S^{*} = \bigcup_{\alpha \in ORD} S^{\alpha}$. Then $S^* = cl_d(A)$. Similarly, we have a characterization of the closure of a subset of a directed space as follows. 

\begin{defn}\rm   \label{IND}
Let $P$ be a set.\ For any convergence class $\mathcal{C} \subseteq \{ (\xi,x): \xi \text{ is a net in }P, x \in P\}$,\ any subset $F \subseteq P$,\ and any ordinal $\alpha \in ORD$,\ we define $F^{\alpha}$ and $F^{*}$ as follows:
\begin{align*}
F^{0}&=F;  \\
F^{\alpha}&=\{ x \in P : \exists \xi \subseteq (\cup_{\beta < \alpha}F^{\beta}),\ (\xi,x) \in \mathcal{C} \}; \\
F^{*}&= \cup_{\alpha \in ORD} F^{\alpha}.
 \end{align*}

\end{defn}

Given any poset $P$ and a subset $A$ of $P$, denote by $ub(A)$ the set of upper bounds of $A$. The cut of $A$ is defined to be $\{x \in P: \forall y \in ub(A), x \leq y\}$, denoted by $A^\delta$.
\begin{prop}\rm  \label{INDU}
Let $(P,\leq)$ be a poset and $\mathcal{C}$ be a convergence class such that $\{ (\{y\},x): x,y \in P, \  x \leq y \} \subseteq \mathcal{C} \subseteq \{ (D,x): D \text{ is a directed subset of }P, \  x \in D^{\delta}\}$.\

\begin{enumerate}
\item[(1)] For any $F \subseteq P$,\ we have $\overline{F} = F^{*}$ in the topological space $(P,\mathcal{C}(P))$.

\item[(2)] The specialization order $\sqsubseteq$ of $(P,\mathcal{C}(P))$ is equal to $\leq$.

\item[(3)] $(P,\mathcal{C}(P))$ is a directed space.
\item[(4)] $F$ is a closed subset of $(P,\mathcal{C}(P))$ iff for any $(D,x) \in \mathcal{C}$ with $D \subseteq F$, $x \in F$.
\end{enumerate}
\end{prop}

\begin{proof}
(1)\ It is easy to see that if $\alpha < \beta$, then $F^{\alpha} \subseteq F^{\beta}$,\ and if $F^{\alpha} = F^{\alpha + 1}$,\ then $F^{*} = F^{\alpha} $.\ For any subset $F$,\ there exists some $\alpha$ such that $F^{\alpha} = F^{\alpha + 1} = F^{*}$ since the cardinal of $F^{*}$ is less than or equal to that of $P$.\ We need only to show that $P \backslash F^{*}$ is open.\ 
For any $(D,x) \in \mathcal{E}$ such that $x \in P\backslash F^{*} $,\ if $D \subseteq F^{*}$,\ then $x \in F^{*}$,\ a contradiction.\ If $D \not\subseteq F^{*}$,\ then $D \cap (P \backslash F^{*}) \neq \emptyset$.\ By $\{(\{y\},x): x,y \in P, \  x \leq y \} \subseteq \mathcal{E}$,\ we know that $P \backslash F^{*} $ is an upper subset of $P$.\ Therefore,\ $D$ is eventually in $ P \backslash F^{*}$,\ i.e.,\ $P\backslash F^{*}$ is open.

(2)\ Given any $y \in P$,\ assume that $(D,x) \in \mathcal{E}$ and $x \in X \backslash\! \da y$.\ Then $D \cap (X \backslash\! \da y) \neq \emptyset$.\ Otherwise,\ $x \in D^{\delta} \subseteq \da y$,\ a contradiction.\ Therefore,\ $P \backslash\! \da y$ is open in $(P,\mathcal{E}(P))$ and then $x \sqsubseteq y$ implies $x \leq y$.\ If $x \leq y$, for any open subset $U$ such that $x \in U$,\ we have $y \in U$ by $ (\{y\},x) \in \mathcal{E}$.\ Thus,\ $x \leq y$ implies $x \sqsubseteq y$. 

(3)\ Let $U \in d(\mathcal{E}(P))$, i.e., for any directed subset $A$ such that $A \to x \in U$ relative to $\mathcal{E}(P)$, $A \cap U \not = \emptyset$. Then, for any $(D,x) \in \mathcal{E}$ such that $x \in U$, since $D \to x$ relative to $\mathcal{E}(P)$, $D \to x$ relative to $d(\mathcal{E}(P))$ by Theorem \ref{convergence}. It follows that $D \cap U \not = \emptyset$ and $D$ is eventually in $U$. Thus, $U$ is an open subset relative to $\mathcal{E}(P)$. $(P,\mathcal{E}(P))$ is a directed space. 

(4)\ By (1), $F$ is closed iff $F = F^*$, iff  $F =F^1$, iff for any $(D,x) \in \mathcal{E}$ with $D \subseteq F$, $x \in F$.
\end{proof}

\begin{rem}\rm
For a $T_0$ space $X$, and any directed subset $D$ of $X$, if $D$ converges to $x \in X$, then $x \in D^\delta$.
\end{rem}

\begin{prop}\rm \label{limit continuity}
Let $Y$ be any topological space, $X$ be a set and $\mathcal{E} \subseteq \{(\xi,x): \xi \text{ is a net in } X, x \in X\}$.
A map $f$ from $(X,\mathcal{E}(X))$ into $Y$  is continuous iff for any $(\xi,x) \in \mathcal{E}$,\ $f(\xi) \rightarrow f(x)$ in $Y$ holds.
\end{prop}

\begin{proof}
The necessity is obvious.\ For sufficiency,\ letting $U$ be an open subset of $Y$,\ we need only to show that $f^{-1}(U)$ is open in $(X,\mathcal{E}(X))$.\ For any $ (\xi,x) \in \mathcal{E}$, and $x \in f^{-1}(U)$,\ we have that $\xi$ is eventually in $f^{-1}(U)$.\ Otherwise,\ $f(\xi)$ is not eventually in $U$,\ a contradiction.
\end{proof}

\section{Representations of free dtop-algebras}

In this section, we first introduce the notion of dtop-algebras based on directed spaces, which are topological generalizations of dcpo-algebras. The existence of the free dtop-algebras over directed spaces with respect to any signature $\Delta$ and a set $\mathcal{E}$ of inequalities was proved by the Adjoint Functor Theorem in \cite{CK2022b}. However, the proof is not constructive.
In this section, we construct the concrete representations of the free dtop-algebras over a directed space via the term algebras. Besides, we show that directed spaces are closely connected to the notion of dcpo presentations defined by Jung, Moshier and Vicker in \cite{AJ2008}. To some extent, we can say that directed spaces are the topological side of the dcpo presentations.

\vskip 3mm

A signature $\Delta = \{\alpha_i\}_I$ consists of a set of operation symbols $\alpha_i$ each being assigned a finite arity $n_i \in \mathbb{N}$. A $\Delta$-algebra  $\langle A,\{f_i\}_I \rangle$ is given by a carrier set $A$ and a set $\{f_i\}_I$ of interpretations of the operation symbols, in the sense that for $i \in I$, $f_i$ is a map from $A^{n_i}$ to $A$. For convenience, we also call each $f_i$ an operation.
A homomorphism between two $\Delta$-algebras $\langle A, \{k_i\}_I \rangle$ and $\langle B ,\{ l_i\}_I\rangle$ is a map $h: A \to B$ which commutes with the operations:
$$\forall i \in I, h(k_i(a_1,\dots,a_{n_i})) = l_i(h(a_1),\dots,h(a_{n_i})).$$

If the carrier set of a $\Delta$-algebra $A$ carries a partial order such that all operations are order preserving, then $A$ is called a (general) partially ordered algebra. If $A$ carries a topology such that all operations $f_i: A^{n_i} \to A$ are separately continuous, then $A$ is called a (general) semi-topological algebra (see \cite{Kei12}). If $A$ is a partially ordered algebra which is a dcpo under the partial oder, and all operations are Scott continuous, then $A$ is called a dcpo-algebra.  

A $ \Delta$-algebra is called a dtop-algebra if the carrier set is equipped with a topology such that it becomes a directed space, and each operation is separately continuous. Note that for a directed space $X$, an operation $f: X^n \to X$ being separately continuous is equivalent to that $f_i$ is continuous from the categorical product of n copies of $X$ to $X$. We let $X^n$ denote the categorical product of n copies of $X$, i.e., $X^n = X \otimes X \otimes \dots \otimes X$. Dcpo-algebras are special dtop-algebras by viewing dpcos as Scott spaces, and dtop-algebras are semi-topological algebras. For dcpo-algebras and directed-algebras, the homomorphisms are required to be continuous. Homomorphisms between dtop-algebras (resp., dcpo-algebras) are also called dtop-homomorphisms (resp., dcpo-homomorphisms).

The term algebra over a set $X$ with respect to a signature $\Delta$ is denoted by $T_\Delta(X)$. It has the universal property that each map from $X$ to $A$, where $\langle A, \{f_i\}_I \rangle$ is a $\Delta$-algebra, can be extended uniquely to a homomorphism $\overline{h}: T_\Delta(X) \to \langle A,\{f_i\}_I \rangle$. Let $V$ be
a fixed countable set of variables.
$T_\Delta(V)$ are used to encode equations. An inequality $\tau_1 \leq \tau_2$ is said to hold in an algebra
$\langle A,\{f_i\}_I \rangle$ if for each map $h: V \to A$ we have $\overline{h}(\tau_1) \leq \overline{h}(\tau_2)$. The pair $\langle \overline{h}(\tau_1),\overline{h}(\tau_2)\rangle$ 
is also called an instance of the inequality $\tau_1 \leq \tau_2$.

We let a pair $\langle \tau_1, \tau_2 \rangle \in  \mathcal{E}  \subseteq T_\Delta(V ) \times  T_\Delta(V)$ stand for the inequality
$\tau_1 \leq \tau_2$. We denote by ${\bf DTop}(\Delta, \mathcal{E})$ the category of all dtop-algebras with signature $\Delta$ that satisfy the inequalities in $\mathcal{E}$, where the morphisms are all the homomorphisms.

\begin{defn}\rm
Let $X$ be a directed space. A dtop-algebra $\langle A,\{f_i\}_I\rangle$ is called the free dtop-algebra over $X$ with respect to $(\Delta,\mathcal{E})$ if $A$ is an object of ${\bf DTop}(\Delta,\mathcal{E})$ and there exists a continuous map $i: X \to A$ such that any continuous map $f: X \to B$, where $\langle B,\{g_i\}_I\rangle$ is an object of ${\bf DTop}(\Delta,\mathcal{E})$, extends uniquely to a homomorphism $\overline{f}: \langle A,\{f_i\}_I\rangle \to \langle B,\{g_i\}_I\rangle$ such that $\overline{f} \circ i =f$.
$$\xymatrix{
  X \ar[r]^{i} \ar[dr]_{\forall f} &  A \ar[d]^{\overline{f}}   & \langle A,\{f_i\}_I\rangle \ar[d]_{\exists!\overline{f}}  \\
 & B & \langle B,\{g_i\}_I\rangle
                     }$$

\end{defn}

\vskip 2mm

Now, we give concrete representations of the free dtop-algebras over a directed space via term algebras.

Let $X$ be a directed space and $T = (\Delta,\mathcal{E})$. For convenience, denote by $X_T$ be the term algebra over $X$, i.e., the set of all terms made from elements of $X$ and operations in $\Delta$. Suppose that $\sqsubseteq$ is the specialization order of $X$. Let $\sqsubseteq_{T}$ be the smallest congruence order on $X_T$ that includes $\sqsubseteq$ (i.e., $a \sqsubseteq_T b$ for all $a \sqsubseteq b$ in $X$, and $w(x_1,\dots,x_n) \sqsubseteq_T w(y_1,\dots,y_n)$ if $x_i \sqsubseteq_T y_i$ for $1 \leq i \leq n$), and satisfies the inequalities in $\mathcal{E}$. 
Given any operation $w$ on $X$ with respect to $T$, let $w_T$ be the corresponding operation on $X_T$ with $w_T(x_1,\dots,x_n) = w(x_1,\dots,x_n)$. Obviously, $w_T$ is monotone.
Define $\mathcal{C}_{T}$ to be the smallest convergence class on $X_T$ such that it includes $DLim(X)$ and $(\{y\},x)$ for all $x \sqsubseteq_T y$ in $X_T$, and such that it is closed under all operations, i.e., if $(D_i,x_i) \in \mathcal{C}_T$, then $(w(D_1 \times \dots \times D_n), w(x_1,\dots,x_n)) \in \mathcal{C}_T$ for any operation $w$. 

\begin{prop}\rm
Let $X$ be a dtop-algebra with respect to $T = (\Delta,\mathcal{E})$. Then $(X_T,\mathcal{C}_T(X_T))$ is the free dtop-algebra over $X$ with respect to $T$.
\end{prop}
\begin{proof}
First, we show that $(X_T,\mathcal{C}_T(X_T))$ is a dtop-algebra with respect to $T$. By Proposition \ref{INDU}, $(X_T,\mathcal{C}_T(X_T))$ is a directed space. Since $\mathcal{C}_T$ is closed under operations, all operations are continuous by Proposition \ref{limit continuity}. Since $w_T(x_1,\dots,x_n) = w(x_1,\dots,x_n)$ for any operation $w_T$, any inequality holds for $X$ is still holds for $X_T$ with respect to $T$. Thus, $(X_T,\mathcal{C}_T(X_T))$ is a dtop-algebra with respect to $T$.

Now, we show that $(X_T,\mathcal{C}_T(X_T))$ is the free dtop-algebra over $X$ with respect to $T$. Let $Y$ be any dtop-algebra with respect to $T$ and $f$ be a continuous map from $X$ to $Y$. The operation on $Y$ corresponding to $w$ on $X$ is denoted by $w_Y$.

$$\xymatrix{
  X \ar[r]^{i} \ar[dr]_{\forall f} & X_T \ar[d]^{\exists ! \overline{f}}  \\
   & Y }$$
Define $\overline{f}: X_T \to Y$ to be

$$
\overline{f}(x)= \begin{cases}
f(x),\quad &x \in X; \\
w_Y(\overline{f}(x_1),\dots,\overline{f}(x_n)),\quad & x = w(x_1,\dots,x_n).
\end{cases} $$

It is easily seen that $f = \overline{f}\circ i$. The commutativity between $\overline{f}$ and any operation $w_T$ follows from the definition of $\overline{f}$. We need only to show that $\overline{f}$ is continuous. Obviously, $\overline{f}$ is monotone. By Proposition \ref{limit continuity}, it is enough to show that for any $(D,x) \in \mathcal{C}_T$, $\overline{f}(D)$ converges to $\overline{f}(x)$. By structure induction, it is true since $\overline{f}$ is equal to $f$ when restricted to $X$ and $\mathcal{C}_T$ is closed under all operations. Thus, $\overline{f}$ is a dtop-homomorphism. Since $f = \overline{f}\circ i$, the uniqueness of $\overline{f}$ follows from the commutativity between $\overline{f}$ and operations $w_T$. 
\end{proof}

\begin{exam}\rm  \label{lower powerspace}
The notion of directed lower powerspaces was defined in the same way as lower powerdomains, and the concrete topological representation of directed lower powerspaces was given in \cite{X2022} as follows.

Let $\Delta$ contain only one binary operation $\vee$, and let $\mathcal{E}$ contain the following equalities.

\begin{enumerate}
\item[(1)] commutativity: $x \vee y = y \vee x $;
\item[(2)] associativity: $(x \vee y) \vee z = x \vee (y \vee z)  $;
\item[(3)] idempotency: $x \vee x = x$.
\item[(4)] inflation: $x \leq x \vee y$.
\end{enumerate}

Then, a dtop-algebra with respect to ${\bf DTop}(\Delta,\mathcal{E})$ is called an inflationary directed semilattice. An inflationary directed semilattice is in fact a directed space with operation join operation as the operation $\vee$ (see \cite{X2022}). The free inflationary directed semilattice over a directed space is called the directed lower powerspace of $X$. It has a concrete topological representation as follows.

	Let\ $X$\ be a directed space.\ Set
$LX = \{\da F :F\subseteq _{fin} X\},  $
where $F\subseteq _{fin} X$\ is an arbitrary nonempty finite subset of \ $X$.\ Define an order\ $\leq_L$\ on\ $LX$\ as follows:
$$\da F_{1}\leq_L \da F_{2}
	\iff \da F_{1}\subseteq \da F_{2}.$$
	Let\ $\mathcal{F}\subseteq LX$\ be a directed set\ (respect to order\ $\leq_L$),\ and $\da F\in LX$.\ Suppose that $F = \{x_1,x_2,\dots,x_n\}$. Define a convergence notation\ $\mathcal{F}\Rightarrow_L \da F$\ as follows:
	\begin{center}
		$\mathcal{F}\Rightarrow _{L}\da F \iff \forall x_i \in  F,\ $\ there exists a directed set\ $D_{i}$\ of\ $X$\ satisfying $D_{i}\subseteq \bigcup \mathcal{F}$\ and\ $D_{i}\rightarrow x_i$.
	\end{center}
A subset\ $\mathcal{U}\subseteq LX$\ is called a\ $\Rightarrow_L$\ convergence open set of\ $LX$\ if and only if for each directed subset\ $\mathcal{F}$\ of\ $LX$\ and\ $\da F\in LX$,\ \ $\mathcal{F}\Rightarrow_{L}\da F\in \mathcal{U}$\ implies\ $\mathcal{F}\cap \mathcal{U}\neq \emptyset$.\ Denote all\ $\Rightarrow_L$\ convergence open sets of\ $LX$\ by\ $O_{\Rightarrow_{L}}(LX)$. Then\ $(LX, O_{\Rightarrow_L}(LX))$ with the set union $\cup$ as operation $\vee$ is the lower powerspace of\ $X$.

The representation of lower powerspaces defined by the above way in \cite{X2022} coincide with ours. Since $\Delta$ contains only one binary operation $\vee$, and $\vee$ satisfies commutativity, associativity and idempotency, the term algebra $X_T$ is restricted to the set of finite subsets of $X$. Let $F_1,F_2$ be two finite subsets of $X$. By the infaltion property, $F_1 \sqsubseteq_T F_2$ iff $\da F_1 \subseteq \da F_2$. Thus, the term algebra $X_T$ is isomorphic to $LX$. 

Supposing that $\mathcal{F} \Rightarrow_U \da F = \da \{x_1,\dots,x_n\}$, by definition, there exist $D_i \subseteq \bigcup \mathcal{F}$, and $D_i \to x_i$ in $X$ for each $x_i \in F$. Since each $D_i$ is contained in the lower set of $\mathcal{F}$, $\mathcal{F}$ converges to $x_i$ for each $i$, and then $\mathcal{F}$ converges to $F$ relative to $\mathcal{C}_T(X_T)$ by the continuity of $\vee$. Thus, $\mathcal{C}_T(X_T)$ is coarser than $O_{\Rightarrow_L}(LX)$. Conversely, it is obvious that $DLim(X)$ is contained in $\Rightarrow_U$ by viewing $x$ as $\da x$ for any $x \in X$. By structure induction, we can get that $\mathcal{C}_T$ is contained in $\Rightarrow_U$. Thus, $\mathcal{C}_T(X_T)$ is finer than $O_{\Rightarrow_L}$. Thus, the $\Rightarrow_U$ convergence defined in \cite{XK2020} generate the same topology as $\mathcal{C}_T$.
\end{exam}
\vskip 3mm

The notion of dcpo presentations was introduced by Jung, Moshier and Vickers in \cite{AJ2008}, for which can generate dcpos. Directed spaces can be viewed as topological dcpo presentations.

\begin{defn} \rm \cite{AJ2008} \begin{enumerate}
\item[(1)] A dcpo presentation consists of a set $P$ of generators, a preorder $\lesssim$ on $P$, and a subset $C$ of $P \times DS(P)$, whose elements $(a,U)$ are called covers and written $a \vartriangleleft U$.
\item[(2)] An order preserving map $f:(P; \lesssim,C) \to D$ from a dcpo presentation to a dcpo $D$ preserves covers if for all $a \vartriangleleft U$ in $C$ it is true that $f(a) \leq \bigvee_{x\in U}f(x)$. 
\item[(3)] A dcpo $\overline{P}$ is freely generated by the dcpo presentation $(P;\lesssim,C)$ if there is a map $\eta:P \to \overline{P}$ that preserves covers, and every map $f$ from $P$ to a dcpo $D$ that preserves covers factors through $\eta$ via a unique Scott-continuous map $\overline{f}:\overline{P} \to D$.
$$\xymatrix{
  \overline{P} \ar[r]^{\exists ! \overline{f}} \ar[d]_{\eta} & D  \\
   P \ar[ur]_{\forall f} }$$
\item[(4)] Let $(P;\lesssim,C)$ be a preorder with cover set. A $C$-ideal $I$ is a subset of $P$ which is downward closed and closed under all covers, to wit, $U \subseteq I$ implies $a \in I$ for all $a \vartriangleleft U$ in $C$.
If $S$ is any subset of $P$ then $\langle S \rangle$ denotes the smallest $C$-ideal containing $S$. The set of all $C$-ideals is denoted by $C$-{\rm Idl}(P).
\end{enumerate}
\end{defn}

Let $P$ be a poset. A set of covers $C$ can also be viewed as a convergence class. Suppose that the cover set $C$ contains all $(x,\{y\})$ for $x \leq y$. Then, by Proposition \ref{INDU}, $\langle S \rangle $ is just the closure of $S$ in the topological space $(P,C(P))$ generated by the cover $C$, which is a directed space, and $C$-Idl(P) is just the topology $C(P)$. Thus, directed spaces can be viewed as topological representations for some special dcpo presentations. By Proposition \ref{limit continuity}, an order preserving map $f:(P; \lesssim,C) \to D$ from a dcpo presentation to a dcpo $D$ preserves covers iff $f:(P,C(P)) \to D$ is continuous. Thus, the freely generated dcpo $\overline{P}$ is exactly the D-completion of $(P,C(P))$.

\vskip 3mm

\section{The D-completion of free dtop-algebras}
In this section, we show that for any dcpo $L$, the D-completion of a free dtop-algebra  over $\Sigma L$ is exactly the free dcpo-algebra over $L$, which reveals the close connection between free dtop-algebras and free dcpo-alegbras. Besides, we show that the dtop-algebra over $\Sigma L$ is continuous iff the dcpo-algebra over $L$ is continuous.

In \cite{Kei12}, it was shown that the separately continuous maps on $T_0$ topological spaces can be naturally lifted to the D-completion of $X$. Let $f:\prod_{i=1}^n X_i \longrightarrow Y$ be a map between topological spaces. Define $f^\gamma: \prod_{i=1}^n \Gamma(X_i) \longrightarrow \Gamma(Y)$ by:
$$f^\gamma(A_1,\dots,A_n) = \overline{f(\prod_{i=1}^n A_i)}$$ 

\begin{lem}\rm \cite{Kei12} \label{lift continuity}
Let $f: \prod_{i=1}^n X_i \longrightarrow  Y$ be separately continuous. Then for arbitrary nonempty subsets $A_i \subseteq X_i$ for $1 \leq i \leq n$ and $B \subseteq Y$:
\begin{enumerate}
\item[(1)] $f(\prod_{i=1}^n \overline{A_i}) 
\subseteq \overline{B}$ whenever $f(\prod_{i=1}^n A_i) \subseteq B$;
\item[(2)]  $\overline{f(\prod_{i=1}^n \overline{A_i})} = \overline{f(\prod_{i=1}^n A_i)} $;
\item[(3)] $f^\gamma: \prod_{i=1}^n  \Gamma(X_i) \longrightarrow  \Gamma(Y)$ is Scott-continuous.
\end{enumerate}
\end{lem}

Restricting $f^\gamma$ on $X^d$, the following statements were proved in \cite{Kei12}.

\begin{prop}\rm \cite{Kei12}
\begin{enumerate}
\item[(1)] A separately continuous function $f : \prod_{i=1}^n X_i \to Y$ extends uniquely to a separately continuous function $f^d: \prod_{i=1}^n X_i^d \to Y^d$. The function $f^d$ is Scott-continuous.

\item[(2)] A homomorphism $f : A \to B$ between semitopological $\Delta$-algebras extends uniquely to a homomorphism $f^d: A^d \to B^d$.

\item[(3)] The basic operations of a topological $\Delta$-algebra $A$ extend uniquely to continuous operations on the D-completion $A^d$
. In this way, the D-completion $A^d$ becomes a algebra of the same signature as $A$, which obeys the same equational and inequational laws as $A$.
\end{enumerate}
\end{prop}

By proposition \ref{D-completion Scott}, the D-completion of a directed space is a Scott space. $X^d = \wt{X}$. Note that $f^d:\prod_{i=1}^n \widetilde{X_i} \to \widetilde{Y}$ is Scott-continuous iff $f^d: \bigotimes_{i=1}^n \Sigma\widetilde{X_i} \to \Sigma\widetilde{Y}$ is a continuous map, iff $f^d$ is separately continuous. Then, we have the following.

\begin{coro}\rm \label{Extension algebra}
\begin{enumerate}
\item[(1)] Let $X_i$ and $Y$ be directed spaces. A continuous map $f : \bigotimes_{i=1}^n X_i \to Y$ extends uniquely to a Scott-continuous map $f^d: \prod_{i=1}^n \wt{X_i} \to \wt{Y}$.

\item[(2)] A continuous homomorphism $f : A \to B$ between dtop-algebras extends uniquely to a continuous homomorphism $f^d: \wt{A} \to \wt{B}$ between dcpos.

\item[(3)] The basic operations of a dtop-algebra $A$ extend uniquely to continuous operations on the dcpo $\wt{A}$, for which has the same signature as $A$, and obeys the same equational and inequational laws as $A$.
\end{enumerate}
\end{coro}

Given any space $X$, let $\eta = \xi_X$, i.e., $\eta: X \to \Gamma(X)$, $\eta(x) = \da x$.

\begin{thm}\rm \label{D-completion free algebra}
If $F(\Sigma L)$ is the free dtop-algebra with respect to $(\Delta,\mathcal{E})$ over a Scott space $\Sigma L$, then $\widetilde{F(\Sigma L)}$ is the free dcpo-algebra with respect to $(\Delta,\mathcal{E})$ over the dcpo $L$.
\end{thm}
\begin{proof}
Suppose that $F(\Sigma L)$ is the free dtop-algebra over $\Sigma L$ with respect to $(\Delta,\mathcal{E})$, by Corollary \ref{Extension algebra}, $\wt{F(\Sigma L)}$ is a dcpo-algebra with respect to $(\Delta,\mathcal{E})$, where the operation on $\wt{F(\Sigma L)}$ corresponding to $w$ on $F(\Sigma L)$ is $w^d$. The dtop-homomorphism $\overline{f}$ is extended to a dcpo-homomorphism $\overline{f}^d$.  Additionally by Theorem \ref{Scott completion}, the following diagram commutes. Thus, $\wt{F(\Sigma L)}$ is the free dcpo-algebra over $Q$ with respect to $(\Delta,\mathcal{E})$.

$$\xymatrix{
  \Sigma L \ar[r]^{i} \ar[dr]_{\forall f} & F(\Sigma L) \ar[r]^{\eta}  \ar[d]^{\exists ! \overline{f}}  & \widetilde{ F(\Sigma L)} \ar[dl]^{\exists ! \overline{f}^d}  \\
   &  \Sigma Q }$$
\end{proof}

\begin{rem} \rm 
Given any space $X$, denote by $F_X$ with $i:X \to F_X$ the free algebra over $X$ with respect to $(\Sigma,\mathcal{E})$ in the context of $\bf Top$, where algebras are semi-topological algebras. 
Recall that functor $\mathcal{D} : {\bf Top} \to {\bf DTop}$ is a coreflector, where $\mathcal{D}(X) = \mathcal{D}X$, $\mathcal{D}(f) = f$ for any object $X$ and any morphism $f$ in $\bf Top$. Let $X$ be a directed space. Given any object $Y$ of ${\bf Top}(\Delta,\mathcal{E})$ and continuous map $f: X \to Y$, since $F_X$ is the free algebra, the following diagram holds.
$$\xymatrix{
  X \ar[r]^{i} \ar[dr]_{\forall f} &  F_X \ar[d]^{\exists ! \overline{f}}     \\
 & Y 
                     }$$
Since $X$ is a directed space, given any space $Z$, map $g: X \to Z$ is continuous iff $g: X \to \mathcal{D}Z$ is continuous. Thus, $i: X \to \mathcal{D}F_X$ is continuous as well. Replace $F_X$ to be $\mathcal{D}F_X$ in the diagram, the diagram still holds. Thus, $\mathcal{D}F_X$ is also the free algebra over $X$ with respect to ${\bf Top}(\Delta,\mathcal{E})$. Since $\mathcal{D}F_X$ is a directed space and the free algebra is unique up to isomorphism, $F_X$ is a directed space. Since $X$ is a directed space, any operation on $X$ with respect to ${\bf Top}(\Delta,\mathcal{E})$ is also an operation with respect to ${\bf DTop}(\Delta,\mathcal{E})$. $F_X$ is also the free dtop-algebra over $X$. Thus, the free algebra over a directed space in the category of topological spaces is equal to the dtop-free algebra over $X$ with respect to the same signature and inequalities. For a dcpo $L$, the free dcpo-algebras over $L$ is equal to the D-completion of the free algebras over $\Sigma L$ in the context of topological spaces as well.

Given any topological space $A$, let $w : A^n \to A$ be any operation in ${\bf Top}(\Delta,\mathcal{E})$. Here, $A^n$ means the topological product, and $w$ is separately continuous. Define $w_D: (\mathcal{D}A)^n \to \mathcal{D}A$ as the same set map of $w$, where $(\mathcal{D}A)^n$ means the categorical product in $\bf DTop$. It is easy to see that $w_D$ is continuous and hence it is an operation on $\mathcal{D}A$ in ${\bf DTop}(\Delta,\mathcal{E})$. 
Thus, we can directly verify that though the functor $\mathcal{D}$, the free dtop-algebra over $\mathcal{D}A$ is exactly $\mathcal{D}F_A$. 

\end{rem}

\vskip 3mm

In \cite[Theorem 4.10]{ZHA2021}, it was shown that for a directed space $X$ whose topology is coarser than the Scott topology, $C(X) \cong C(\widetilde{X})$. In fact, this is also true for all directed  spaces. Given any directed space $X$ and any subset $A$ of $\widetilde{X}$, denote by $cl_{\widetilde{X}}(A)$ the Scott closure of $A$ in $\widetilde{X}$, and denote by $cl_d(A)$ the d-closure of $A$ in $\widetilde{X}$.

\begin{prop}\rm \label{topology equal}
Let $X$ be a directed space. Then $C(X) \cong C(\widetilde{X})$.
\end{prop}
\begin{proof}
Suppose that $X$ is a directed space. Define 
$$ i: C(X) \to C(\widetilde{X}), i(A) = cl_{\widetilde{X}} (\Psi(A)) =  cl_{\widetilde{X}} (\{\da a: a \in A\}), \forall A \in C(X), $$
and $$\bigcup:  C(\widetilde{X}) \to C(X), \bigcup(\mathcal{A}) = \bigcup \mathcal{A}, \forall \mathcal{A} \in C(\widetilde{X}).$$

1. Obviously, $i$ is well-defined and monotone, and $\bigcup$ is monotone. Given any $\mathcal{A} \in  C(\widetilde{X})$ and directed subset $D \subseteq \bigcup \mathcal{A}$, $\{\da d : d \in D\}$ is a directed subset of $\mathcal{A}$. Then $\bigvee_{\widetilde{X}} \{\da d: d \in D\} = cl_X (D) \in \mathcal{A}$. Thus, $cl_X(D) \subseteq \bigcup \mathcal{A}$. It follows that $\bigcup \mathcal{A}$ is closed in $X$. Thus, $\bigcup$ is well-defined.

2. $i\circ \bigcup = 1_{C(\widetilde{X})}$. Given any $\mathcal{A} \in C(\widetilde{X})$, $i\circ \bigcup (\mathcal{A}) = cl_{\widetilde{X}}(\{\da x : x \in \bigcup \mathcal{A}\}) = cl_{\widetilde{X}}(\mathcal{A} \bigcap \Psi(X))$. Since in $\widetilde{X}$, the d-closure of $\Psi(X)$ is $\widetilde{X}$ and $\mathcal{A}$ is Scott closed, it is a lower set and d-closed. We have $\mathcal{A} \subseteq \widetilde{X} = cl_d(\Psi(X)) = cl_d(\Psi(X) \cap \mathcal{A}) \cup cl_d(\Psi(X) \cap \mathcal{A}^c)$, and $cl_d(\Psi(X)\cap \mathcal{A}^c) \subseteq \mathcal{A}^c$. Since $\mathcal{A}^c$ is d-closed, then $\mathcal{A} \subseteq cl_d(\Psi(X) \cap \mathcal{A})$. Thus, $\mathcal{A} = cl_d ( \mathcal{A} \cap  \Psi(X)) \subseteq cl_{\widetilde{X}}(\mathcal{A} \cap  \Psi(X)) \subseteq \mathcal{A}$. Thus, $i \circ \bigcup(\mathcal{A}) = \mathcal{A}$.

3. $\bigcup \circ i = 1_{C(X)}$. Given any $A \in C(X)$, $\bigcup \circ i(A) = \bigcup cl_{\widetilde{X}}(\{\da a : a \in A\})$. Obviously, $A \subseteq \bigcup \circ i(A)$. Given any $x \in \bigcup cl_{\widetilde{X}}(\{\da a : a \in A\})$, there exits some $B \in cl_{\widetilde{X}}(\{\da a : a \in A\})$ such that $x \in B$. Since $B \subseteq cl_{X} A = A$, we have $x \in A$. Thus, $\bigcup \circ i(A) = A$.

By the above, we conclude that $C(X) \cong C(\widetilde{X})$.
\end{proof}

\begin{coro} \rm
Let $X$ be a directed space. Then $\eta: X \to \widetilde{X},\ \eta(x) = \da x$ is a topological embedding.
\end{coro}
\begin{proof}
By proposition \ref{Scott completion}, $\eta$ is continuous. Given any closed subset $A$ of $X$, $\eta(A) = \{\da a: a \in A\} \subseteq \eta(X) \cap cl_{\widetilde{X}}(\{\da a : a \in A\})$. Conversely, suppose that $\da x \in cl_{\widetilde{X}}(\{\da a:a\in A\})$. Since $A$ is a closed subset of $X$, given any $F \in cl_{\widetilde{X}}(\{\da a:a\in A\})$, $F \subseteq A$. Thus, $\da x \subseteq A$, and then $x \in A$. It follows that $\eta(A) = \{\da a: a \in A\} = \eta(X) \cap cl_{\widetilde{X}}(\{\da a : a \in A\})$. Thus, $\eta$ is a topological embedding.
\end{proof}

\begin{coro} \rm
Let $L$ be a dcpo. Then the free dtop-algebra over $\Sigma L$ with respect to $(\Sigma,\mathcal{E})$ is a subspace of the free dcpo-algebra over $L$ with respect to $(\Sigma,\mathcal{E})$ up to isomorphism.
\end{coro}

The continuity of a dcpo can be extended to a directed space. Let $X$ be a directed space. For any $x,y \in X$, define $x \ll y$ iff for each directed subset $D$ of $X$ such that $D \to y$, $D \cap \ua x \not = \emptyset$. If $x \ll x$, then $x$ is called a compact element.

\begin{defn}\rm  \cite{XK2022,ZHA2021}
$X$ is called a continuous space if it is a directed space such that $\DDa x = \{a \in X: a \ll x\}$ is directed and converges to $x$ for every $x \in X$. $X$ is called an algebraic space if it is a directed space such that $comp(x) = \{a \in X: a \ll a \ \& \ a \leq x\}$ is directed and converges to $x$ for every $x \in X$.
\end{defn}
A $T_0$ topological space $X$ is called a c-space if for any $x \in U \in \mathcal{O}(X)$, there exists a $y \in U $ such that $x \in (\ua y)^\circ$. $X$ is called a b-space if for any $x \in U \in \mathcal{O}(X)$, there exists a $y \in U $ such that $x \in (\ua y)^\circ = \ua y$. Continuous spaces are exactly all the c-spaces, and algebraic spaces are exactly all the b-spaces.

\begin{prop}\rm \cite{CKL2022} \label{continuity cor}
A directed space $X$ is continuous (resp., algebraic) iff $C(X)$ is a continuous (resp., algebraic) dcpo.
\end{prop}

Combining Proposition \ref{topology equal} and Proposition \ref{continuity cor}, we have the following.

\begin{coro} \rm 
Let $X$ be a directed space. Then $X$ is continuous (resp., algebraic) iff $\widetilde{X}$ is a continuous (resp., algebraic) dcpo.
\end{coro}

\begin{coro} \rm 
Let $L$ be a dcpo. Then the free dtop-algebra over $\Sigma L$ is a continuous space iff the corresponding free dcpo-algebra over $L$ is a continuous dcpo.
\end{coro}

In \cite{CK2022b}, it was shown that the free dtop-algebra over a continuous (resp., algebraic) space is continuous (resp., algebraic). Then, the following statement is a direct corollary.

\begin{coro}\rm  \cite{AJ94}
Let $L$ be a continuous (resp., algebraic) dcpo. Then the free dcpo-algebra over $L$ is continuous. 
\end{coro}

\section{Applications}

Among those free dcpo-algebras over dcpos, there are three kinds of classical ones: the upper, lower and convex powerdomains (see \cite{AJ94,Hec91}). They are used to model the denotational semantics for non-deterministic functional programming languages.  In \cite{X2022}, the concrete topological representations of directed upper, lower and convex powerspaces were given. In \cite{ZHA2017}, a concrete representation of the D-completion of any space was given via tapered closed subsets. As applications of Theorem \ref{D-completion free algebra}, we can gain the topological representation of the three kinds of powerdomains via the D-completion of the directed powerspace uniformly.

\subsection{Convex powerdomains}
Convex powerdomains are also called Plotkin powerdomains, which are free dcpo semilattices over dcpos. For a countably based or coherent continuous domain $L$, the convex powerdomain $L$ has a simple topological representation consisting of lens (see \cite{CON03}). However, there is no topological representation for convex powerdomains of general dcpos. In the following, we give the representation via directed convex powerspaces. First, we introduce the notion of directed convex powerspaces and give their topological representations.

\vskip 3mm
\
Let $\Delta$ contain only one binary operation $+$, and let $\mathcal{E}$ contain the following equalities.

\begin{enumerate}
\item[(1)] commutativity: $x + y = y +x $;
\item[(2)] associativity: $(x + y) + z = x + (y + z)  $;
\item[(3)] idempotency: $x + x = x$.
\end{enumerate}

Then, a dtop-algebra with respect to $(\Delta,\mathcal{E})$ is called a directed semilattice. The free directed semilattice over a directed space is called the directed convex powerspace of $X$. A dcpo-algebra with respect to $(\Delta,\mathcal{E})$ is called a dcpo semilattice. Dcpo semilattices 
can be viewed as special directed semilattices by viewing dcpos as directed spaces endowed with the Scott topology. Directed convex powerspaces have a concrete topological representation as follows (see \cite{XK2022}).

\begin{defn}\rm  \cite{XK2022}
Let $X$ be a directed space. Define	$\widehat{F}=({\da} F,{\ua} F)$ and $CX=\{\widehat{F} :F\subseteq _{fin} X\},$
where $F\subseteq _{fin} X$ denote that $F$ is a nonempty finite subset of $X$. Define order $\leq_C$ on $CX$~as follows:

	$$\widehat{F_{1}}\leq_C \widehat{F_{2}}
	\iff {\da} F_{1}\subseteq {\da} F_{2} ~\text{and}~{\ua F}_{2}\subseteq {\ua} F_{1}.$$
	
	Suppose that $\mathcal{D}=\{\widehat{F_{i}}\}_{i\in I}\subseteq CX$ is a directed subset of $CX$ (with order$\leq_C$). For convenience, denote by $$\pi_{1}\widehat{F}={\da} F,\pi_{2}\widehat{F}={\ua} F,$$
	and $$\pi_1\mathcal D=\{{\da} F_i\}_i,\pi_{2}\mathcal D=\{{\ua} F_i\}_i.$$
	
	Suppose that $\widehat{F}\in PX, F=\{a_1,\dots,a_n\}$. Define $\mathcal{D}\Rightarrow_C \widehat{F}$ as follows.
	
	$\mathcal{D}\Rightarrow _{C}\widehat{F}  \iff $ there exists finite directed subset $D_{1},\dots,D_{n}$ of $X$ satisfy the folloing conditions:
	\begin{enumerate}
		\item[(1)]$D_i\subseteq\bigcup\limits_{i\in I}\pi_1\mathcal D,i=1,\dots,n$;
		\item[(2)] $\forall i=1,\dots,n$, $D_i\rightarrow a_i$;
		\item[(3)] $\forall (d_{1},\dots,d_{n})\in\prod\limits_{i=1}^{n}D_{i}, \exists \widehat{F^{'}}\in\mathcal{D}$ such that $\pi_2\widehat{F^{'}}\subseteq {\ua}(d_{1},\dots,d_{n})$.
	\end{enumerate}

A subset $\mathcal{U}\subseteq PX$ is called a $\Rightarrow_C$ convergence open subset of $PX$ if for any directed subset $\mathcal{D}=\{\widehat{F_{i}}\}_{i\in I}$ of $PX$ and $\widehat{F}\in PX$, 
	$$\mathcal{D}\Rightarrow_{P}\widehat{F}\in \mathcal{U}\Rightarrow\mathcal{D}\cap \mathcal{U}\neq \emptyset.$$ 
Denote by $O_{\Rightarrow_{C}}(PX)$ the set of all $\Rightarrow_C$ convergence open subsets of $PX$.\\ 
\end{defn} 

Denote by $\mathcal{P}X = (PX,O_{\Rightarrow_C}(PX))$. Define $+: \mathcal{P}X \otimes \mathcal{P}X \to \mathcal{P}X$ as follows: $\forall \widehat{F_{1}}, \widehat{F_{2}} \in \mathcal{P}X$, $\widehat{F_{1}} + \widehat{F_{2}} = \widehat{F_1 \cup F_2}$.

\begin{prop}\rm  \cite{X2022}
Let $X$ be a directed space. Then, $\mathcal{P}X$ with operation $+$ is the directed convex powerspace of $X$ up to isomorphism.
\end{prop}

The directed convex powerspace of a directed space $X$ is unique up to isomorphism. For a dcpo $L$, denote by $\mathcal{P}L$ the directed convex poewrspace of $\Sigma L$. Then, by Theorem \ref{D-completion free algebra}, we gain a topological representation of convex powerdomains as follows.

\begin{coro}\rm
The convex powerdomain of a dcpo $L$ is equal to $\wt{\mathcal{P}L}$ with the operation $+^d$ up to isomorphism.
\end{coro}





\subsection{Upper powerdomains}

Let $\Delta$ contain only one binary operation $\wedge$, and let $\mathcal{E}$ contain the following equalities and inequalities.

\begin{enumerate}
\item[(1)] commutativity: $x \wedge y = y \wedge x $;
\item[(2)] associativity: $(x \wedge y) \wedge z = x \wedge (y \wedge z)  $;
\item[(3)] idempotency: $x \wedge x = x$.
\item[(4)] deflation: $x \geq x \wedge y$.
\end{enumerate}

Then the dtop-algebra with respect to $(\Delta,\mathcal{E})$ is called a deflationary directed semilattice. A deflationary directed semilattice is in fact a directed space with the meet operation as $\wedge$ \cite{X2022}. The free deflationary directed semilattice over a directed space $X$ is called the directed upper powerspace of $X$. It has a topological representation as follows.

\begin{defn} \rm
Let\ $X$\ be a directed space.\ Denote
$$UX=\{\uparrow F :F\subseteq _{fin} X\},$$
here,\ $F\subseteq _{fin} X$\ is an arbitrary nonempty finite subset of\ $X$.\ Define an order\ $\leq_U$\ on\ $UX$\ :
$$\uparrow F_{1}\leq_U \uparrow F_{2}
	\iff \uparrow F_{2}\subseteq \uparrow F_{1}.$$
	Let\ $\mathcal{F}\subseteq UX$\ be a directed set(respect to order\ $\leq_U$)\ and\ $\ua F\in UX$.\ Define a convergence notation\ $\mathcal{F}\Rightarrow_{U}\ua F\iff$\ there exists finite directed sets\ $D_{1},\dots,D_{n}\subseteq X$\ such that
\begin{enumerate}
	\item $F\cap \lim D_{i}\neq \emptyset,\ \forall D_{i}$;
	\item $F\subseteq \bigcup\limits_{i=1}^{n}\lim D_{i}$;
	\item $\forall (d_{1},\dots,d_{n})\in \prod\limits_{i=1}^{n}D_{i}$,\ there exists some\ $\uparrow F^{'}\in \mathcal{F}$,\ such that\ $\uparrow F^{'}\subseteq \bigcup\limits_{i=1}^{n}\uparrow d_{i}$.
\end{enumerate}
A subset\ $\mathcal{V}\subseteq UX$\ is called a\ $\Rightarrow_U$\ convergence open set of\ $UX$\ if and only if for each directed subset\ $\mathcal{F}$\ of\ $UX$\ and\ $\ua F\in UX$,\ $\mathcal{F}\Rightarrow_{U}\uparrow F\in \mathcal{V}$\ implies\ $\mathcal{F}\cap \mathcal{V}\neq \emptyset$.\ Denote all\ $\Rightarrow_U$\ convergence open set of\ $UX$\ by\ $O_{\Rightarrow_{U}}(UX)$. 
\end{defn}

\begin{thm}\rm \cite{X2022} 
Let $X$ be a directed space. Then, $(UX,O_{\Rightarrow_U}(UX))$ with the set union $\cup$ as the operation $\wedge$ is the upper powerspace of $X$.
\end{thm}

For a dcpo $L$, denote by $\mathcal{U}L$ the directed upper powerspace of $\Sigma L$. By Theorem \ref{D-completion free algebra}, we gain the topological representation of the upper powerdomain as follows.

\begin{prop}\rm 
Let $L$ be a dcpo. Then $\wt{\mathcal{U}L}$ is the upper powerdomain of $L$ up to isomorphism.
\end{prop}

It is well known that for a continuous dcpo $L$, the upper powerdomain of $L$ has a simple topological representation $\mathcal{Q}(L)$. 

\begin{thm}\rm {\rm \cite[Theorem IV-8.10]{CON03}}
The Smyth powerdomain over a continuous domain $L$ may be realized
as the set $\mathcal{Q}(L)$, ordered by reverse inclusion. The embedding $j$ of $L$ into
$\mathcal{Q}(L)$ is given by $j(x) = \ua x$.
\end{thm} 

Now, we investigate this property from the viewpoint of the D-completion of directed spaces, and try to explain why the topological representation of upper powerdomains of general dcpos is hard to describe.

The following are some basic properties of $\mathcal{Q}(P)$ for a continuous dcpo $P$.

\begin{lem}\rm \cite{Hec92,HK13} \label{SP}
Let $P$ be a continuous dcpo. Then $\mathcal{Q}(P)$ is a continuous semilattice and $\forall A,B \in \mathcal{Q}(P)$,
\begin{enumerate}
\item[(1)] $A \ll_Q B$ ($\ll_Q$ denote the way-below relation in $\mathcal{Q}(P)$) iff there exists a finite subset $F \subseteq_f P$ such that $B \subseteq (\ua F)^\circ \subseteq {\ua F} \subseteq A$.
\item[(2)] $\{{\ua F}: \emptyset \not = F \subseteq_f P\}$ is a basis of $\mathcal{Q}(P)$.
\item[(3)] $A \wedge B = A \cup B$.
\end{enumerate}
\end{lem}

Let $L$ be a dcpo. Given any $K \in \mathcal{Q}(L)$, define $\mathcal{F}_K = \{\ua F: K \subseteq \ua F \ \& \ F \subseteq_f L\}$. 

\begin{lem}\rm \label{closure equal}
Let $L$ be a dcpo. For each $K \in \mathcal{Q}(L)$, $\mathcal{F}_K$ is a closed subset of $\mathcal{U}L$.
\end{lem}
\begin{proof}
Given any $K \in \mathcal{Q}(L)$, it is easy to see that $\mathcal{F}_K$ is a lower subset of $\mathcal{U}L$. Suppose that $\mathcal{G}$ is a directed subset of $\mathcal{F}_K$ and $\mathcal{G} \Rightarrow_U \ua G = \ua \{a_1,a_2,\dots,a_n\}$. By Proposition \ref{INDU}, we need only to show that $\ua G \in \mathcal{F}_K$, i.e., $K \subseteq \ua G$. By the definition of $\Rightarrow_U$ convergence, there exist finite directed subsets $D_1,\dots,D_k$ of $X$ such that:

(1) For each $a_i$, there is a $1\leq j_i \leq k$ such that $D_{j_i} \to a_i$.

(2) For each $D_j$, there is a $1\leq i_j \leq n$ such that $D_j \to a_{i_j}$.

(3) For each $t = (x_1,\dots,x_k) \in \prod_{1\leq i \leq k} D_i$, there exists some $\ua G_t 
\in \mathcal{G}$ such that $\ua G_t \subseteq \ua  \{x_1,\dots,x_n\}$. 

Since $K \subseteq \ua G_i$ for each $\ua G_i \in \mathcal{G}$,  $K \subseteq \bigcap \mathcal{G} \subseteq \bigcap \{\ua \{x_1,\dots,x_k\}: (x_1,\dots,x_k) \in  \prod_{1\leq i \leq k} D_i \}$. Suppose that there exists an $x \in K$ such that $x \not \in \ua G$. Then, $L \backslash \da x$ is an Scott open subset of $L$ that contains $\{a_1,\dots,a_n\}$. By (1) and (2), there exists some $w = (y_1,\dots,y_k) \in \prod_{1\leq i \leq k} D_i$ and $\{y_1,\dots,y_k\} \in L \backslash \da x$, i.e., $x \not \in \ua \{y_1,\dots,y_k\}$. It follows that there exists some $G_w \in \mathcal{G}$ such that $x \not \in G_w$, a contradiction. Thus, $K \subseteq \ua G$. \end{proof}

\begin{prop}\rm \label{upper equal}
Let $L$ be a continuous domain. Given any directed subset $\mathcal{G}$ of $\mathcal{Q}(L)$, then $\bigcap \mathcal{G} \in \mathcal{Q}(L)$ and $ \mathcal{F}_{\bigcap \mathcal{G}} =  cl_{\mathcal{U}L}(\bigcup_{K \in \mathcal{G}}\mathcal{F}_K) = \bigvee_{C(\mathcal{U}L)} \{\mathcal{F}_K: K \in \mathcal{G}\}$.
\end{prop}
\begin{proof}
Since $L$ is continuous, $\Sigma L$ is locally compact sober. Thus, given any directed subset $\mathcal{G}$ of $\mathcal{Q}(L)$, $\bigcap \mathcal{G} \in \mathcal{Q}(L)$. Given any  $\ua F = \ua \{x_1,\dots,x_n\} \in \mathcal{F}_{\bigcap \mathcal{G}}$, i.e., $\bigcap \mathcal{G} \subseteq \ua F$, define $$\mathcal{G}_F= \{\ua \{a_1,\dots,a_n\}: a_i \ll x_i, \forall 1 \leq i \leq n\}.$$
Then, $\mathcal{G}_F$ is a directed subset of $\mathcal{U}L$  and  $\mathcal{G}_F \Rightarrow_U \ua F$. To show $\ua F \in cl_{\mathcal{U}L}(\bigcup_{K \in \mathcal{F}}\mathcal{F}_K)$, we need only to show that for each $\ua \{a_1,\dots,a_n\} \in \mathcal{G}_F$, $\ua \{a_1,\dots,a_n\} \in \bigcup_{K \in \mathcal{F}}\mathcal{F}_K$. Since $\bigcap \mathcal{G} \subseteq \ua \{x_1,\dots,x_n\} \subseteq \UUa\{a_1,\dots,a_n\} \subseteq \ua\{a_1,\dots,a_n\}$, by the sobriety of $L$, there exists some $\ua G \in \mathcal{G}$ such that $\ua G \subseteq \UUa\{a_1,\dots,a_n\}$. Thus, $\ua \{a_1,\dots,a_n\} \in \mathcal{F}_{\ua G}$. It follows that $\ua F  \in cl_{\mathcal{U}L}(\bigcup_{K \in \mathcal{F}}\mathcal{F}_K) $, i.e., $ \mathcal{F}_{\bigcap \mathcal{G}} \subseteq  cl_{\mathcal{U}L}(\bigcup_{K \in \mathcal{F}}\mathcal{F}_K)$.

By Lemma \ref{closure equal}, $ \mathcal{F}_{\bigcap \mathcal{G}}$ is a closed subset of $\mathcal{U}L$. Since $\bigcup_{K \in \mathcal{G}} \mathcal{F}_K \subseteq \mathcal{F}_{\bigcap \mathcal{G}}$, then $ cl_{\mathcal{U}L}(\bigcup_{K \in \mathcal{G}}\mathcal{F}_K) \subseteq \mathcal{F}_{\bigcap \mathcal{G}}$.
\end{proof}

\vskip 3mm

The $\Rightarrow_U$ convergence satisfies the condition of Proposition \ref{INDU}. Recall that for a dcpo $P$, and a subset $A$ of $P$, let $S^{0} = A$,   
$S^{\alpha + 1} = \{x : x = \bigvee D, D$  is a directed subset of $S_{\alpha} \} $, $S^\beta = \bigcup_{\alpha < \beta} S^\alpha$ for limit ordinal $\beta$, and  $S^{*} = \bigcup_{\alpha \in ORD} S^{\alpha}$. Then, $S^* = cl_d(A)$ \cite{Kei09}.

\begin{thm}\rm
Let $L$ be a continuous domain. Then $\widetilde{\mathcal{U}L} = \{\mathcal{F}_K: K \in \mathcal{Q}(L)\} \cong \mathcal{Q}(L)$.
\end{thm}
\begin{proof}
By Lemma \ref{closure equal}, for each $K \in \mathcal{Q}(L)$, $\mathcal{F}_K = \{\ua F : K \subseteq \ua F,\ F \subseteq_f L\}$  is a closed subset of $\mathcal{U}L$. Suppose $K_1,K_2 \in \mathcal{Q}(L)$ and $K_1 \not = K_2$. Without loss of generality, suppose that $x \in K_1, x \not \in K_2$. Then $K_2 \in L \backslash \da x$. By the continuity of $L$, there exists a finite subset $F \subseteq L \backslash \da x$ with $K_2 \subseteq \ua F$ and $K_1 \not \subseteq \ua F$. Thus, $\mathcal{Q}(L)$ is order isomorphic to $\{\mathcal{F}_K:K \in \mathcal{Q}(L)\} \subseteq C(\mathcal{U}L)$.  We need only to show that $\widetilde{\mathcal{U}L} = \{\mathcal{F}_K:K \in \mathcal{Q}(L)\}$.

Since $L$ is continuous, $\mathcal{Q}(L)$ has a basis of $\{\ua F: F \subseteq_f L\}$ by Lemma \ref{SP}. Given any $K \in \mathcal{Q}(L)$, define $\mathcal{F}^*_K = \{\ua F: F \subseteq_f L \ \& \ K \subseteq (\ua F)^\circ \}$. Then $\mathcal{F}^*_K$ is a directed subset of $\mathcal{Q}(L)$ and $\bigcap \mathcal{F}^*_K = K$. By Proposition \ref{upper equal}, $\mathcal{F}_K = \mathcal{F}_{\bigcap \mathcal{F}^*_K} = \bigvee_{C(\mathcal{U}L)} \{\mathcal{F}_B:B \in \mathcal{F}^*_F\}$. Let $S^0 = \Psi(\mathcal{U}L)$,  
$S^{\alpha + 1} = \{x : x = \bigvee D, D$  is a directed subset of $S_{\alpha} \} $, $S^\beta = \bigcup_{\alpha < \beta} S^\alpha$ for limit ordinal $\beta$, and  $S^{*} = \bigcup_{\alpha \in ORD} S^{\alpha}$. Then $S^* = \wt{\mathcal{U}L}$.  Since for each $B \in \mathcal{F}^*_F$, $\mathcal{F}_B = \big\downarrow_{\mathcal{U}L} B$, we have $\mathcal{F}_B \in S^0$. Thus, $\mathcal{F}_K \in S^1 \subseteq \widetilde{\mathcal{U}L}$ for every $K \in \mathcal{Q}(L)$.

Conversely, we know $S^0 \subseteq \{\mathcal{F}_K: K \in \mathcal{Q}(L)\}$. Assume that for any $\alpha < \beta$, $S^\alpha \subseteq \{\mathcal{F}_K: K \in \mathcal{Q}(L)\}$. Suppose $A \in S^\beta$ and $A = \bigvee_{C(\mathcal{U}L)}\mathcal{A}$ for directed subset $\mathcal{A} \subseteq \bigcup_{\alpha < \beta} S^\alpha$. Then by Proposition \ref{upper equal}, $A \in \{\mathcal{F}_K: K \in \mathcal{Q}(L)\}$. It follows that $\widetilde{\mathcal{U}(L)} = S^1 = S^* = \{\mathcal{F}_K: K \in \mathcal{Q}(L)\}$.
\end{proof}

From the above proof, we see that for a continuous dcpo $L$, $\mathcal{Q}(L)$ is isomorphic to $S^1 = \{\mathcal{F}_K: K \in \mathcal{Q}(L)\}$, which is also equal to $\wt{\mathcal{U}L}$. Thus, the upper powerdomain of $L$ has $\mathcal{Q}(L)$ as a simple topological representation. For general dcpos, $S^1$ is not equal to the upper powerdomain of $L$ by mapping $K \in \mathcal{U}L$ to $\mathcal{F}_K$. Thus, the upper powerdomain of an arbitrary dcpo $L$ may not be isomorphic to $\mathcal{Q}(L)$.

\vskip 3mm
\subsection{Lower powerdomains}
The case of lower powerdomains is simpler since for an arbitrary dcpo $P$, the lower powerdomain of $P$ has a topological representation $\Gamma(P)$. In the following, we analysis this in the viewpoint of D-completion of directed spaces.

The notion of lower powerspaces was displayed in Example \ref{lower powerspace}.
The lower powerspace of a directed space is unique up to isomorphism,\ we denote the lower powerspace of each directed space\ $X$ by\ $\mathcal{L}X=(LX,\ \cup)$. For a dcpo $P$, we denote by $\mathcal{L}P$ the directed lower powerspace of $\Sigma P$. Since the lower powerdomain of a dcpo $P$ has the simple topological representation $\Gamma(P)$, by Theorem \ref{D-completion free algebra}, we see that $\wt{\mathcal{L}P} \cong \Gamma(P) $. Now, we give a direct proof of $\widetilde{\mathcal{L}P} \cong \Gamma(P)$ from the viewpoint of D-completion.

\vskip 3mm

\begin{prop}\rm \label{equal trans lower}
Let $P$ be a dcpo. Given any directed subset $\mathcal{F} = \{ \da F_i\}_{i \in I}$ of $\mathcal{L}P$,  we have $cl_{\mathcal{L}P}(\mathcal{F}) = \{\da F: F \subseteq_f cl_{\Sigma P}(\bigcup \mathcal{F})\}$.
\end{prop}
\begin{proof}
Consider $S^0 = \mathcal{F} =  \{\da F_i\}_{i\in I}$, the subset of $\mathcal{L}P$. Let $S^*$ be the set generated by the $\Rightarrow_{L}$ convergence (see Definition \ref{IND}). Then $S^* = cl_{\mathcal{L}P}(\mathcal{F})$ by Proposition \ref{INDU}. To show $cl_{\mathcal{L}P}(\mathcal{F}) \subseteq \{\da F: F \subseteq_f cl_{\Sigma P}(\bigcup \mathcal{F})\}$, we need only to show that for each ordinal $\alpha$, $S^\alpha \subseteq \{\da F: F \subseteq_f cl_{\Sigma P}(\bigcup \mathcal{F})\}$. When $\alpha = 0$, it is obviously true. Given any ordinal $\beta$, suppose that for all $\alpha < \beta$, the inclusion holds, and $\da G \in S^\beta$. Then there exists some a directed subset $\mathcal{G}$ of $\bigcup_{\alpha < \beta} S^\alpha$ such that $\mathcal{G} \Rightarrow_L \da G$, that is, given any $a \in G$, there exists a directed subset $G_a \subseteq \bigcup \mathcal{G} \subseteq \bigcup(\bigcup_{\alpha < \beta} S^\alpha)$ such that $G_a \to a$ in $\Sigma P$. Since $G_a \subseteq \bigcup(\bigcup_{\alpha < \beta} S^\alpha) \subseteq cl_{\Sigma P}(\bigcup\mathcal{F})$, $a \in cl_{\Sigma P}(\bigcup\mathcal{F})$. Thus, $G \subseteq_f cl_{\Sigma P}(\bigcup \mathcal{F})$. By transfinite induction, for all $\alpha$, $S^\alpha \subseteq \{\da F: F \subseteq_f cl_{\Sigma P}(\bigcup \mathcal{F})\}$. It follows that for  $cl_{\mathcal{L}P}(\mathcal{F}) \subseteq \{\da F: F \subseteq_f cl_{\Sigma P}(\bigcup \mathcal{F})\}$.

Conversely, let $F^0 = \bigcup \mathcal{F}$. Then $F^0$ is a lower subset of $X$. Let $F^{\alpha + 1} = \{x : x \leq \bigvee D, D$  is a directed subset of $S^{\alpha} \} $, $F^\beta = \bigcup_{\alpha < \beta} S^\alpha$ for limit ordinal $\beta$, and  $F^{*} = \bigcup_{\alpha \in ORD} F^{\alpha}$. Then $F^* = cl_{\Sigma P}(\bigcup \mathcal{F})$. To show $\{\da F: F \subseteq_f cl_{\Sigma P}(\bigcup \mathcal{F})\} \subseteq cl_{\mathcal{L}P}(\mathcal{F})$, we need only to show that for each ordinal $\alpha$ and a subset $A \subseteq_f F^\alpha$, $\da A \in cl_{\mathcal{L}P}(\mathcal{F})$.

When $\alpha = 0$, given any subset $A \subseteq_f F^0 = \bigcup \mathcal{F}$, there exist subsets $\da F_1, \da F_2,\dots,\da F_n \in \mathcal{F}$ such that $A \subseteq \bigcup_{i=1}^n \da F_i$. Since $\mathcal{F}$ is directed, there is some $\da F^\prime \in \mathcal{F}$ such that $\bigcup_{i=1}^n\da F_i \subseteq \da F^\prime$. Thus, $A \subseteq \da F^\prime$. It follows that $\{\da F^\prime\} \Rightarrow_L \da A$, and then $\da A \in cl_{\mathcal{L}P}(\mathcal{F})$. 

Given any ordinal $\beta$, suppose that for all $\alpha < \beta$, given any subset $A \subseteq_f F^\alpha$, $\da A \in cl_{\mathcal{L}P}(\mathcal{F})$.
If $\beta$ is a limit ordinal, then given any $F \subseteq_f F^\beta$, there exists some $\alpha_0 < \beta$ such that $F \subseteq F^{\alpha_0}$. Thus, $\da F \in cl_{\mathcal{L}P}(\mathcal{F})$.  If $\beta = \alpha +1$ for some $\alpha$, given any $F = \{x_1,\dots,x_n\} \subseteq_f F^\beta$, there exists some directed subset $D_i \subseteq F^\alpha$ such that $D_i \to x_i$ with respect to $\Sigma P$ for each $1 \leq i \leq n$. Then, $\mathcal{H} = \{\da \{x_1,\dots,x_n \}: (x_1,\dots,x_n) \in \prod_{i=1}^n D_i\}$ is a directed subset of $\mathcal{L}P$, and $\mathcal{H} \Rightarrow_L \da F$ by definition of $\Rightarrow_L$ convergence. Since each element of $\mathcal{H}$ is in $\mathcal{F}$, $\da F \in cl_{\mathcal{L}P}(\mathcal{F})$. By transfinite induction, for each ordinal $\alpha$ and a subset $A \subseteq_f F^\alpha$, $\da A \in cl_{\mathcal{L}P}(\mathcal{F})$. 
\end{proof}

\begin{thm} \rm
Let $P$ be a dcpo. Then $\widetilde{\mathcal{L}P}$ is equal to $\Gamma(P)$ up to isomorphism. Every $K \in \Gamma(P)$ is mapped to $\mathcal{F}_K = \{\da F: F \subseteq_f K\} \in \wt{\mathcal{L}P}$.
\end{thm}
\begin{proof}
Given any $K \in \Gamma(P)$, denote by $\mathcal{F}_K = \{\da F: F \subseteq_f K\}$. Then $\mathcal{F}_K$ is a directed subset of $\Gamma (P)$ and $K = \bigcup \mathcal{F}_K = \bigvee_{\Gamma(P)} \mathcal{F}_K$. $\mathcal{F}_K$ is also a directed subset of $\mathcal{L}P$. Hence, $\mathcal{F}^\prime_K = \{\big\downarrow_{\mathcal{L}P} \da F: \da F \in \mathcal{F}_K\}$ is a directed subset of $\Psi(\mathcal{L}P)$ and $C(\mathcal{L}P)$. We have

$$\bigvee_{C(\mathcal{L}P)}  \mathcal{F}^\prime_K =  cl_{\mathcal{L}P}(\bigcup_{F \subseteq_f K} \big\downarrow_{\mathcal{L}P} \da F) 
 = cl_{\mathcal{L}P}(\{\da F: F \subseteq_f K\}).$$
 By Proposition \ref{equal trans lower}, $\da G \in cl_{\mathcal{L}P}(\{\da F: F \subseteq_f K\})$ iff $\da G \subseteq cl_{\Sigma P}(\bigcup \mathcal{F}_K) = K \in \Gamma(P)$. Thus
 $$\bigvee_{C(\mathcal{L}P)}  \mathcal{F}^\prime_K = \{\da F: F \subseteq_f K\} = \mathcal{F}_K. $$ 
Let $S^0 = \Psi(\mathcal{L}P)$, $S^{\alpha + 1} = \{x : x = \bigvee D, D$  is a directed subset of $S_{\alpha} \} $, $S^\beta = \bigcup_{\alpha < \beta} S^\alpha$ for limit ordinal $\beta$, and  $S^{*} = \bigcup_{\alpha \in ORD} S^{\alpha}$. Then $S^* = \wt{\mathcal{L}P}$. It follows that 
$\mathcal{F}_K$ is a closed subset of $\mathcal{L}P$ and $\mathcal{F}_K \in S^1 \subseteq S^* = \wt{\mathcal{L}P}$. 

Conversely, given any directed subset $\mathcal{F} = \{\big\downarrow_{\mathcal{L}P} \da F_i \}_{i \in I}$ of $S^0$. Then
$$\bigvee_{C(\mathcal{L}P)}  \mathcal{F} = cl_{\mathcal{L}P}(\bigcup \mathcal{F})
= cl_{\mathcal{L}P}(\{\da F_i: i \in I \}). $$
By Proposition \ref{equal trans lower}, $\da G \in cl_{\mathcal{L}P}(\{\da F_i: i \in I \})$ iff  $\da G \subseteq cl_{\Sigma P}(\bigcup \{F_i: i\in I\})$. Thus, $$\bigvee_{C(\mathcal{L}P)}  \mathcal{F}  = \{\da F: F\subseteq_f cl_{\Sigma P}(\bigcup\{F_i:i \in I\})\}.$$
Thus, $S^1 = \{\mathcal{F}_K : K \in \Gamma(P)\}$.
Let $\mathcal{D} = \{\mathcal{F}_{K_i} : K_i \in \Gamma(P), i \in I\}$ be a directed subset of $S^1$. Then $\bigcup \mathcal{D}$ is a directed subset of $\mathcal{L}P$, and $$\bigvee_{C(\mathcal{L}P)} \mathcal{D} = cl_{\mathcal{L}P}(\bigcup \mathcal{D}) = \{\da F \in \mathcal{L}P: \da F \subseteq cl_{\Sigma P}(\bigcup \{F: \da F \subseteq \bigcup(\bigcup \mathcal{D}) \}) \} \in S^1.$$
Thus, $S^2 = S^1$, then, $S^* = S^1$, and there is a one-to-one correspondence between $\Gamma(P)$ and $S^*$ by mapping any $K \in \Gamma(P)$ to $\mathcal{F}_K$.
\end{proof}

\end{document}